\documentclass[prl,twocolumn,superscriptaddress]{revtex4-2}
\usepackage{dcolumn}
\usepackage{bm}
\usepackage{graphicx}
\usepackage{amsmath}
\usepackage{amssymb}
\usepackage[ansinew]{inputenc}
\usepackage{epstopdf}
\usepackage{multirow}
\usepackage[normalem]{ulem}
\newcommand{\eis}{Eu$_5$In$_2$Sb$_6$}
\newcommand{\nsize}{\fontsize{11pt}{11pt}\selectfont}

\usepackage{xcolor}
\usepackage[OT4]{fontenc}
\usepackage[urlcolor=blue,colorlinks=true,citecolor=blue,linkcolor=blue,
            bookmarks=false,hypertexnames=false]{hyperref}

\begin{document}
\title{Thermodynamic evidence for polaron stabilization inside the
antiferromagnetic order of Eu$_5$In$_2$Sb$_6$}

\author{H. Dawczak-D\k{e}bicki}
\affiliation{Max Planck Institute for Chemical Physics of Solids, 01187
Dresden, Germany}

\author{M. Victoria Ale Crivillero}
\affiliation{Max Planck Institute for Chemical Physics of Solids, 01187
Dresden, Germany}

\author{M. S. Cook}
\affiliation{Los Alamos National Laboratory, Los Alamos, NM 87545, USA}

\author{S. M. Thomas}
\affiliation{Los Alamos National Laboratory, Los Alamos, NM 87545, USA}

\author{Priscila F. S. Rosa}
\affiliation{Los Alamos National Laboratory, Los Alamos, NM 87545, USA}

\author{J. M\"{u}ller}
\affiliation{Institute of Physics, Goethe-University Frankfurt, 60438
Frankfurt (M), Germany}

\author{U. K. R\"{o}{\ss}ler}
 \affiliation{IFW Dresden, Helmholtzstra{\ss}e 20, 01069 Dresden, Germany}

\author{P. Schlottmann}
\affiliation{Department of Physics, Florida State University, Tallahassee,
Florida 32306, USA}

\author{S. Wirth}
\affiliation{Max Planck Institute for Chemical Physics of Solids, 01187
Dresden, Germany}
\date{\today}\maketitle

\textbf{
Materials exhibiting electronic inhomogeneities at the nanometer scale have
enormous potential for applications. Magnetic polarons are one such type of
inhomogeneity which link the electronic, magnetic and lattice degrees of
freedom in correlated matter and often give rise to colossal magnetoresistance.
Here, we investigate single crystals of Eu$_5$In$_2$Sb$_6$ by thermal expansion
and magnetostriction along different crystallographic directions. These data
provide compelling evidence for the formation of magnetic polarons in
Eu$_5$In$_2$Sb$_6$ well above the magnetic ordering temperature. More
specifically, our results are consistent with anisotropic polarons with varying
extent along the different crystallographic directions. A crossover revealed
within the magnetically ordered phase can be associated with a surprising
stabilization of ferromagnetic polarons within the global antiferromagnetic
order upon decreasing temperature. These findings make Eu$_5$In$_2$Sb$_6$ a
rare example of such coexisting and competing magnetic orders and, importantly,
shed new light on colossal magnetoresistive behavior beyond
manganites.}\\

\noindent {\large \textbf{Introduction}}\\
Materials in which the electrical and magnetic properties are intertwined are
of considerable interest from a fundamental point of view as well as for
applications, specifically in spintronics \cite{kho97,zie01}. Here, correlated
materials are of particular significance because of their often inherent
coupling of lattice, charge, orbital and/or spin degrees of freedom. As a
consequence of these multiple interactions electronically inhomogeneous states
can arise \cite{dag05}. Moreover, these multiple interactions may give rise to
large responses upon small perturbations, a desired scenario for possible
applications. Prime examples of electronically correlated, spatially
inhomogeneous materials are the cuprate superconductors with high transition
temperatures \cite{eme90,kei15} or manganites exhibiting colossal
magnetoresistance (CMR) effects \cite{coe99}.

Eu-based compounds are often prone to coupled charge and spin degrees of
freedom: A strong exchange interaction between charge carriers and the
localized spin of the Eu$^{2+}$ ions \cite{kas68,die94} may result in a
localization of these charge carriers at Eu$^{2+}$ sites. The resulting
quasiparticle made up of the localized charge carriers and the Eu moment is
referred to as a magnetic polaron \cite{mol07,yak10} and exemplifies the
interplay between electrical and magnetic properties, the hallmark of
spintronics \cite{wol01,zut04}. A prerequisite for polaron formation is a low
enough carrier concentration, which assures polarons remain localized and do
not populate the materials volume. In this respect, Zintl phases \cite{zin39}
emerge as a promising playground. In particular, Eu$_5$In$_2$Sb$_6$ has
been recently reported \cite{ros20,ale22,ale23} as a narrow-gap semiconductor
\cite{par02,sub16} whose properties are consistent with polaron formation.
Notably, this material is a magnetic counterpart to Ba$_5$In$_2$Sb$_6$
\cite{cor88}, which crystallizes in the non-symmorphic space group $Pbma$
(No.\ 55) and has been predicted to display topological surface states
\cite{par13,wie18}. Hence, Eu$_5$In$_2$Sb$_6$ may combine topologically
nontrivial surface states with magnetism and polaron formation, albeit only
the latter two properties have been experimentally established so far
\cite{ros20,ale22,ale23}. Also, a remarkably large negative magnetoresistance,
comparable to the best CMR manganites, was observed in Eu$_5$In$_2$Sb$_6$
\cite{ros20,ale23}. The price to be paid for this potentially interesting
combination of properties is the complex orthorhombic structure of
Eu$_5$In$_2$Sb$_6$ with three crystallographically different Eu sites
\cite{par02}, resulting in a complicated magnetic phase diagram with at least
two antiferromagnetic phases ($T_{\rm N1} \approx$ 14.1~K, $T_{\rm N2} \approx$
7.2~K) and entangled magnetization processes \cite{ros20,ale22,mor23,rah23}. It
must also be pointed out that despite an orbital angular momentum of $L = 0$ of
the Eu$^{2+}$ ions in this compound, strong anisotropies of the transport
properties and the magnetic susceptibility in the magnetically ordered phase
have been observed indicating competing magnetic interactions. In particular,
the low symmetry of the Eu sublattices allows for the existence of
Dzyaloshinskii-Moriya interactions and spin canting \cite{ale22}. Thus, the
magnetic structure of Eu$_5$In$_2$Sb$_6$ is highly complex \cite{ale22,mor23,
rah23} and not yet fully resolved in all detail.

In an effort to shed light on the intriguing properties of Eu$_5$In$_2$Sb$_6$
we conducted thermal expansion and magnetostriction measurements along the
different crystallographic directions of single crystals. The results provide
further evidence for the formation of non-spherical polarons far above the
N\'{e}el temperature $T_{\rm N}$. These polarons give rise to an extraordinarily
rich and anisotropic magnetoresistive behavior. Within the antiferromagnetically
ordered phase, our results are in excellent agreement with the established
phase diagrams. Additionally, however, these dilatation data reveal a crossover
which is associated with the coexistence of ferromagnetic (fm) polarons and
antiferromagnetic (afm) long-range order. Surprisingly, these polarons become
more stable upon lowering the temperature. This crossover could not be traced
by any other, i.e.\ magnetic, transport or specific heat measurement, a fact
\begin{figure}[t]
\centering
\includegraphics[width=8.6cm]{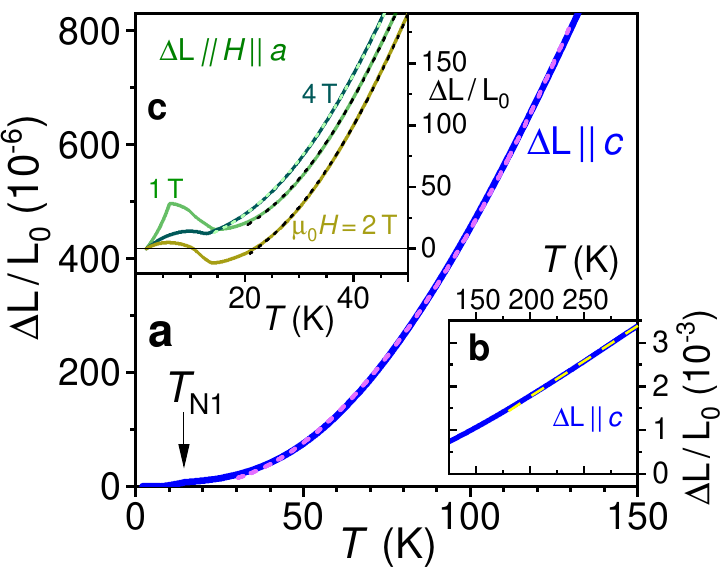}
\caption{\textbf{Fits of thermal expansion data.} {\bf a} $\Delta
{\rm L}(T)/$L$_0$ can very well be fitted to a model containing electronic and
phononic contributions \cite{muk96} for 35~K$\,\lesssim T\lesssim\,$130~K and
{\bf b} is linear in $T$ above 175~K, respectively. {\bf c}: Fits to the model
\cite{muk96} for $\Delta {\rm L} (T,H)/$L$_0$ measured along the $a$ axis
and at selected fields (same data as in Fig.\ \ref{te-ms}\textbf{a}, units of
$10^{-6}$). All dashed lines are fit results.}  \label{fits}
\end{figure}
that points to magnetoelastic coupling effects as its origin and that
underscores the remarkable sensitivity of expansion measurements, establishing
them as pivotal tool to unveil subtle phenomena.\\

\noindent {\large \textbf{Results}}\\
\noindent \textbf{Paramagnetic regime}\\
We start off by considering the paramagnetic regime before discussing the
magnetically much more complex afm low-temperature regime. Thermal expansion
data up to room temperature for measurements along the $c$ axis are presented
in Fig.\ \ref{fits}. For $T \gtrsim$ 175~K, the relative length change
$\Delta$L$(T) /$L$_0$ increases linearly with $T$ within experimental
resolution (Fig.\ \ref{fits}{\bf b}), as expected for a material with
\begin{figure}[t]
\centering
\includegraphics[width=8.4cm]{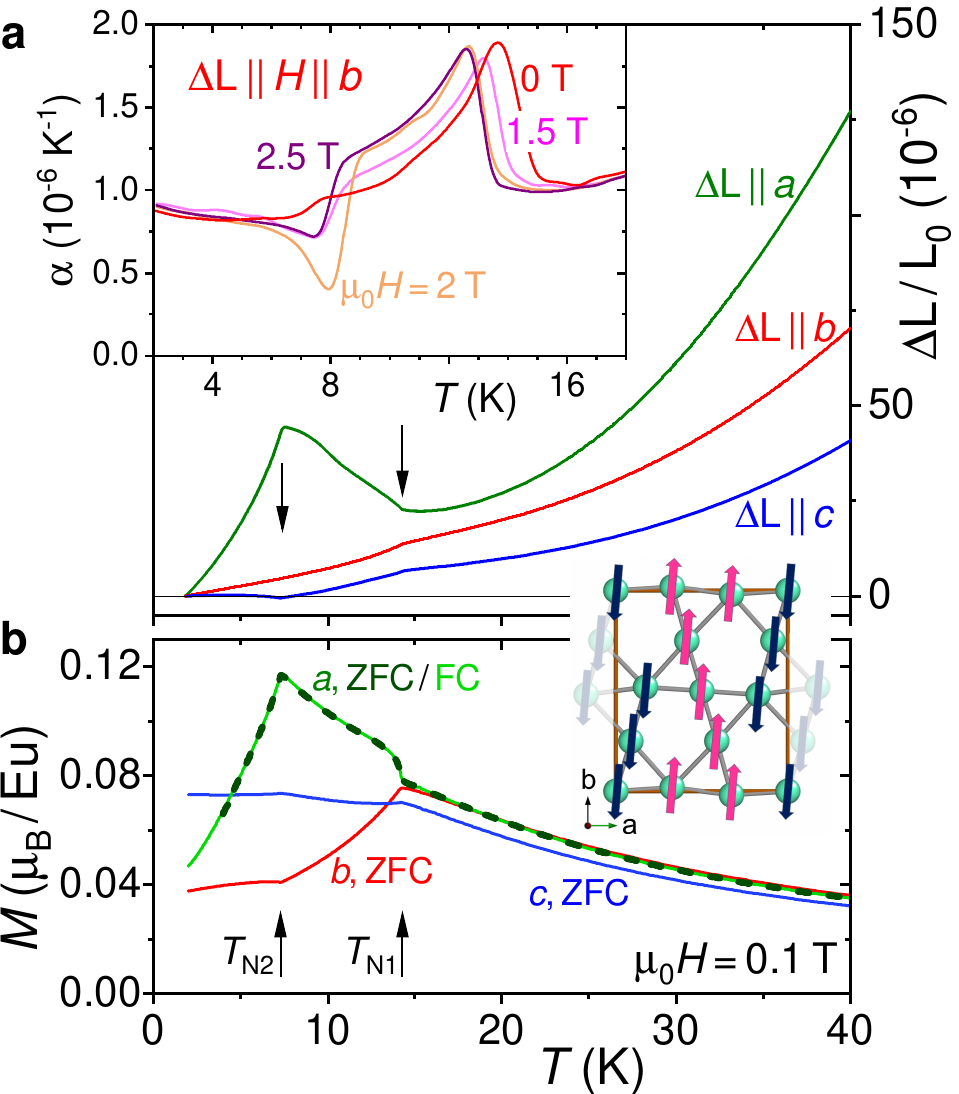}
\caption{\textbf{Comparison of thermal expansion and magnetization.}
{\bf a} Relative length change $\Delta {\rm L}(T) / $L$_0$ measured along
the three main crystallographic directions of Eu$_5$In$_2$Sb$_6$. Inset:
Low-temperature thermal expansion coefficient $\alpha (T, H)$ along the $b$
axis and for representative magnetic fields $\mu_0 H =$ 0, 1.5, 2.0, 2.5~T.
The feature near 8~K evolves non-monotonically with $H$. {\bf b} Zero-field
cooled (ZFC) magnetization of Eu$_5$In$_2$Sb$_6$ measured in $\mu_0 H =$
0.1~T for the three directions. For comparison, the field-cooled (FC) curve
along the $a$ axis is also shown. Arrows mark $T_{\rm N1}$ and $T_{\rm N2}$,
respectively. Sketch: spin configuration just below $T_{\rm N1}$ in phase
I, as suggested in \cite{ale22}.}  \label{aniso}
\end{figure}
a Debye temperature $\theta_{\rm D} \approx$ 197~K \cite{ale23}. Here, L$_0$
is the sample length at a reference temperature. Between 35~K $\lesssim T
\lesssim$ 130~K, Fig.\ \ref{fits}{\bf a}, $\Delta {\rm L}(T) /$L$_0$
can be well described by a model in which electronic and phononic contributions
are considered \cite{muk96,lin02}, see Supplementary Note 2 and Supplementary
Table I. For simplicity, we relied on our specific heat results \cite{ale23}
and conducted the fits with $\theta_{\rm D}=$ 197~K and without Einstein modes.
For measurements along the $a$ axis, the fits hold down to about 25~K, and with
an applied field of 4~T even to slightly lower $T$, cf.\ Fig.\
\ref{fits}{\bf c}. We note that using different fitting procedures does not
change our findings qualitatively (see Supplementary Figures 2 and 3).

Since thermodynamic, magnetic and electronic transport properties of
Eu$_5$In$_2$Sb$_6$ all indicate a clear impact of magnetic polarons below 40 --
50~K \cite{ros20,gho22,sou22,ale23}, the deviations of $\Delta
{\rm L}(T)/$L$_0$ from the modelled $T$-behavior can safely be associated with
such polarons. In Eu$_5$In$_2$Sb$_6$, these nanoscale clusters are likely
anisotropic in shape with the shortest direction along $c$ and fm alignment of
the Eu$^{2+}$ moments within the polaron in the $ab$ plane \cite{ale22,sou22,
gho22}. It is well established that magnetic polarons can induce lattice
distortions \cite{det97,man14,fra21}. In Fig.\ \ref{aniso}{\bf a}, we focus on
$T \le$ 40~K and present the relative length changes $\Delta {\rm L}(T)/$L$_0$
as measured along the three main crystallographic directions of
Eu$_5$In$_2$Sb$_6$. Clearly, there are large anisotropies observed, as already
seen in magnetic, transport and piezoresistive measurements \cite{ros20,gho22,
ale23} and treated by density functional theory calculation \cite{ale22}. For
all directions, however, a kink is observed at $\approx$14.3~K in zero magnetic
field, a temperature associated with $T_{\rm N1}$ \cite{ros20,ale23}. The
transition near $T_{\rm N2} \approx$ 7.2~K is most pronounced for $\Delta
{\rm L}^{100}(T) / $L$_0$ along the $a$ axis. In fact, the strong length
increase with increasing $T$ below 7.2~K and the subsequent decrease up to
14.3~K mimics very nicely the strongly $T$-dependent magnetization $M(T)$
measured along the $a$ axis, as shown in Fig.\ \ref{aniso}{\bf b}. Similarly,
thermal expansion and magnetization exhibit the smallest change with
\begin{figure}[t]
\centering
\includegraphics[width=8.8cm]{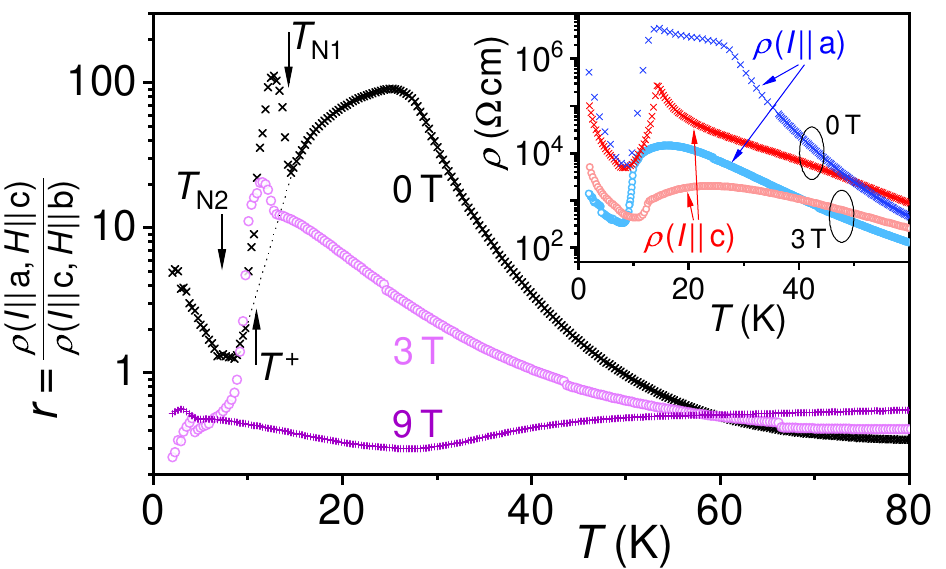}
\caption{\textbf{Analysis of resistivities.} Ratio $r$ of resistivities $\rho$
measured along $I \parallel a$, {\bf H} $\parallel c$ versus $I \parallel c$,
{\bf H} $\parallel b$ for fields $\mu_0 H =$ 0~T ($\times$), 3~T ($\circ$) and
9~T ($+$). The dotted line is a guide to the eye. Inset: individual $\rho$
values; $I \parallel a$ bluish, $I \parallel c$ reddish for 0~T ($\times$) and
3~T ($\circ$). $T^+$ marks the crossover into an inhomogeneous, polaronic
state upon lowering $T$, here shown for $H = 0$.}  \label{resist}
\end{figure}
temperature for measurements along the $c$ axis. Here we emphasize again that
$L = 0$ for Eu$^{2+}$ and hence, crystalline electric field (CEF) effects are
negligible to first order. This implies that magnetic interactions not only
determine the anisotropic magnetization behavior in Eu$_5$In$_2$Sb$_6$ but also
its thermal expansion in magnetic fields and magnetostrictive behavior
\cite{lin02,doe05}.

We now attempt to interpret the resistive behavior of Eu$_5$In$_2$Sb$_6$,
Fig.\ \ref{resist}. In order to minimize dependencies on specific samples
on the one hand, and focus on the anisotropy of the polarons on the
other hand, we consider the ratio of resistivities along different
directions, $r = \rho_{I\parallel a} / \rho_{I\parallel c}$, where $\rho_{I
\parallel a}$ and $\rho_{I\parallel c}$ are the resistivities measured for
$I\parallel a$ ({\bf H} $\parallel c$) and $I\parallel c$ ({\bf H} $\parallel
b$), respectively \cite{ale23}. The infinite [In$_2$Sb$_6]^{10-}$ ribbons of
the Eu$_5$In$_2$Sb$_6$ structure along $c$ support preferred charge transport
along this direction \cite{par02} upon lowering $T$ below 50~K, and $r(T)$
increases strongly at zero field. This increase of $r(T)$ is intercepted below
30~K by the interaction of the polarons which are more extended in the $ab$
plane. This results in a shoulder of $\rho_{I \parallel a}$ and a pronounced
maximum in $r(T)$. The sharp increase of $r(T)$ below $T =$ 14.4~K reflects
the onset of afm ordering at $T_{\rm N1}$.

In a magnetic field of 3~T parallel to $c$, the polarons grow in size along
$c$ and hence, abate their anisotropy, such that the differences in $\rho_{I
\parallel a}$ and $\rho_{I \parallel c}$ (for fields along $c$ and $b$,
respectively) are much less pronounced; likely the maximum in $r(T)$ is then
shifted close to $T_{\rm N1}$. In the nearly field-polarized state at 9~T
(saturation of magnetization is achieved near 10~T at 2~K \cite{ale23}), the
fm polarons have grown further, overlapped and likely merged into their
\begin{figure}[b]
\centering
\includegraphics[width=8.9cm]{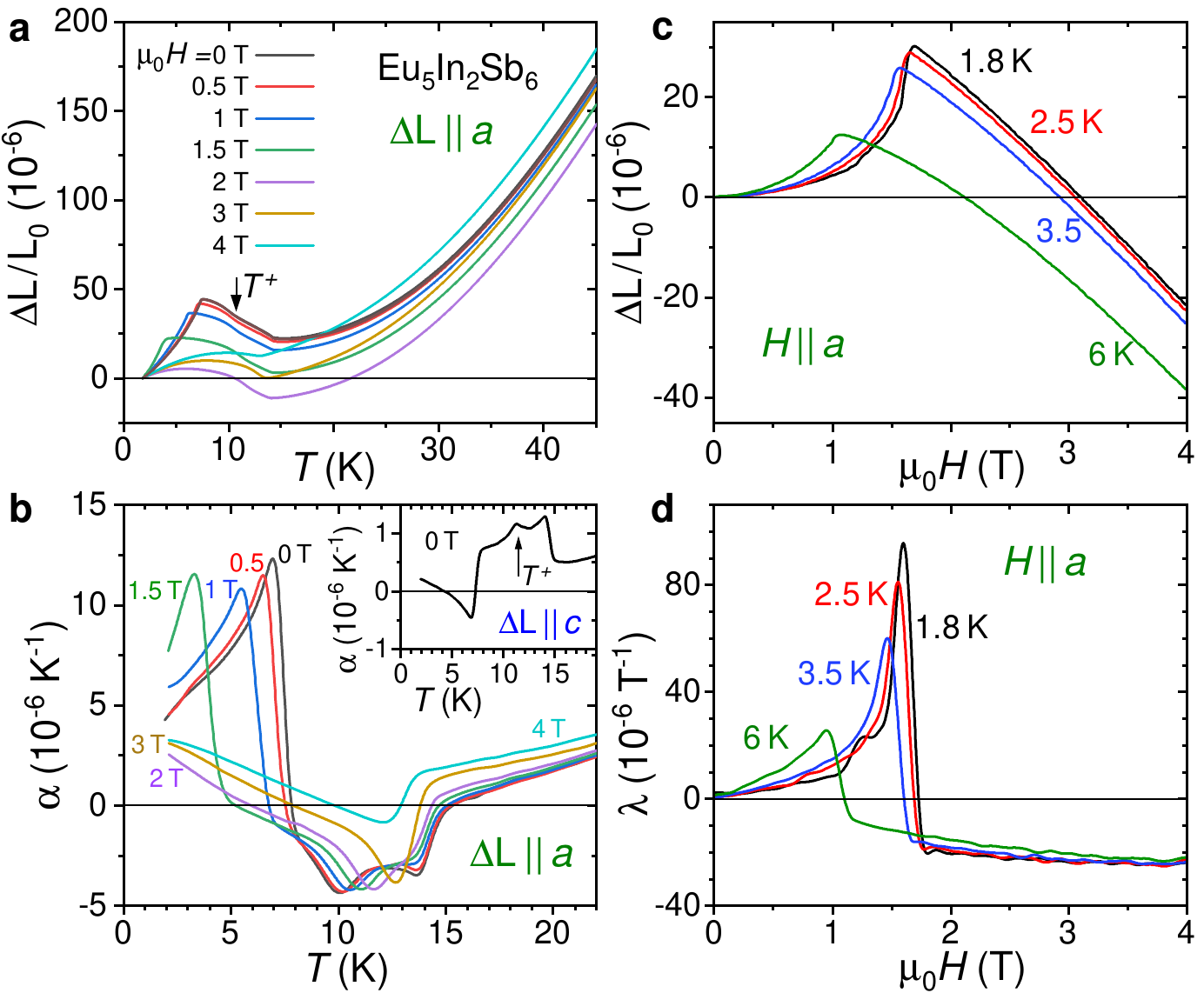}
\caption{\textbf{Detailed thermal expansion and magnetostriction along $a$
axis.} {\bf a} Length change $\Delta{\rm L}(T) / $L$_0$ and {\bf b} thermal
expansion coefficient $\alpha (T)$ measured at different magnetic fields $H$
along the $a$ axis. Inset: $\alpha (T, H=0)$ along the $c$ axis. {\bf c}
Magnetostriction $\Delta {\rm L}(H) / $L$_0$ and {\bf d} magnetostriction
coefficient $\lambda$ for {\bf H} $\parallel a$ at different $T$.}
\label{te-ms}  \end{figure}
surroundings. In consequence, only a weak temperature dependence of $r(T)$ is
observed.\\

\noindent \textbf{Antiferromagnetically ordered regime}\\
Considering the formation of polarons well above the afm ordered phase as
being established we now investigate the fate of these polarons upon cooling to
below $T_{\rm N1}$. The transitions at $T_{\rm N1}$ and $T_{\rm N2}$ can easily
\begin{figure*}[t]
\centering
\includegraphics[width=17.8cm]{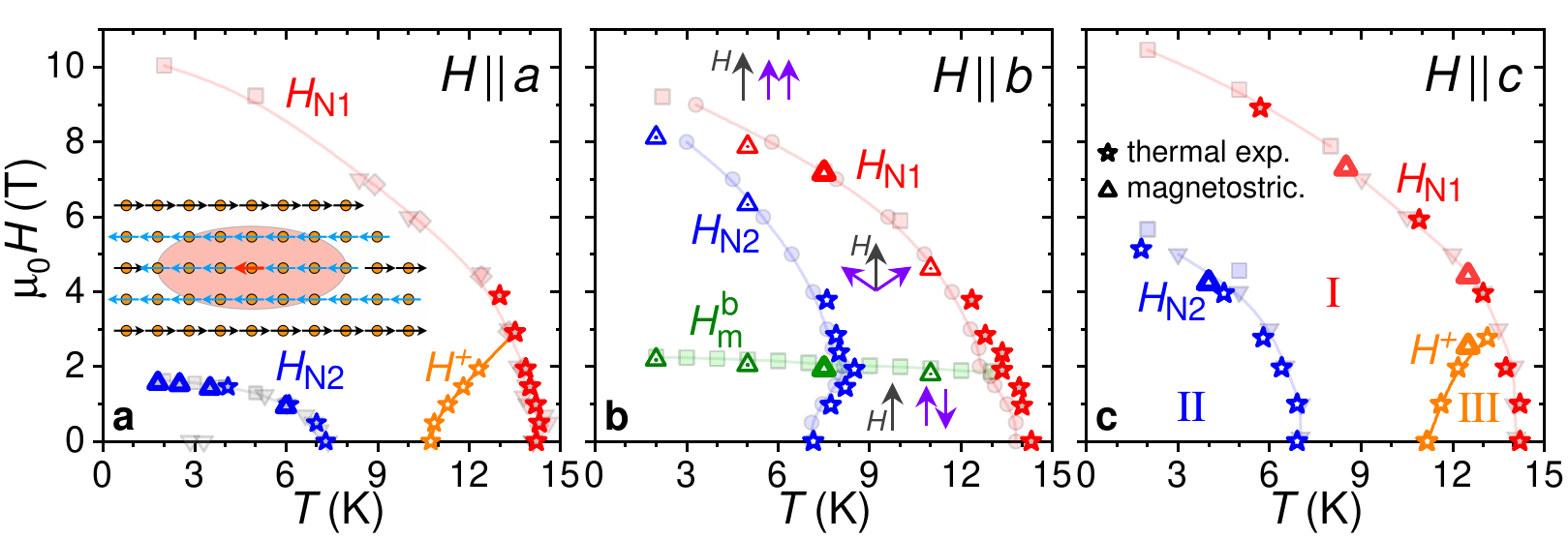}
\caption{\textbf{$H$--$T$ phase diagrams of Eu$_5$In$_2$Sb$_6$} for the three
main crystallographic directions: {\bf a} along $a$ axis, {\bf b} along $b$
and {\bf c} along $c$. The faint lines and data points, reproduced
\cite{ale23} for direct comparison, result from specific heat and magnetic
measurements. Thermal expansion and magnetostriction results are indicated by
$\star$ and $\vartriangle$, respectively (dotted triangles for {\bf H}
$\parallel b$ indicate results obtained with a different setup, see
Supplementary Note 4 and Supplementary Figure 5). $H_{\rm m}^b$ marks a
metamagnetic behavior. The three regions I, II and III between $H_{\rm N1}$
and $H_{\rm N2}$, below $H_{\rm N2}$, and between $H_{\rm N1}$ and $H^+$,
respectively, are marked in {\bf c}. Here, $H_{\rm N1}$ and $H_{\rm N2}$ mark
antiferromagnetic (afm) phase transitions while $H^+$ signals a crossover. In
{\bf a}, an anisotropic ferromagnetic polaron at zero field is sketched within
afm global order, a situation expected in region I. The polarons' shortest
dimension is expected along the crystallographic $c$ direction and the Eu
moments are aligned in the $ab$ plane.}  \label{phasedia}
\end{figure*}
be recognized in the dilatation data along the different crystallographic
directions, Fig.\ \ref{aniso} and Fig.\ \ref{te-ms}, particularly from the
coefficients of thermal expansion $\alpha (T) = (1 / {\rm L}_0)(d{\rm L} /d T)$
and magnetostriction $\lambda =$ (1/L$_0)\, \partial {\rm L} / \partial H$. In
Fig.\ \ref{phasedia} we include these results into the $T$--$H$ phase
diagrams obtained from magnetic, specific heat and transport measurements
(faint data points and lines from Ale Crivillero \textit{et al.}\cite{ale23})
for the three main crystallographic directions. In order to allow for such a
comparison the samples' demagnetizing effects were taken into consideration.
$H_{\rm N1}$ and $H_{\rm N2}$ describe the field dependencies of $T_{\rm N1}$
and $T_{\rm N2}$, respectively. Clearly, our dilatation data trace the
transitions at $H_{\rm N1}$ (red) and  $H_{\rm N2}$ (blue) very well. This
emphasizes again that the effects observed in thermal expansion and
magnetostriction are primarily related to the magnetic structure of
Eu$_5$In$_2$Sb$_6$. Further support for such an assignment comes from the
magnetostriction measurement $\Delta {\rm L} (H) / $L$_0$ at $T =$ 7.5~K along
the crystallographic $b$ direction which exhibits a jump at $\mu_0 H_{\rm m}^b
\approx$ 2~T, red and orange data in Fig.\ \ref{ms001}{\bf b}. Here,
$H_{\rm m}^b$ marks the metamagnetic behavior associated with a spin flop
\cite{ale23} as sketched in Fig.\ \ref{phasedia}{\bf b}.\\

\noindent \textbf{Analysis of magnetostrictive behavior}\\
As clearly seen in $\alpha (T,H)$ along the $a$ and the $c$ axis these data
reveal another, field-dependent feature just below $T_{\rm N1}$ denoted as
$H^+$ in Fig.\ \ref{phasedia} (orange data). In the following, we inspect
this feature by considering the magnetostriction for {\bf H} $\parallel c$
within the three regions marked in Fig.\ \ref{phasedia}{\bf c}. Note that
$c$ is the magnetically hard direction and applying {\bf H} $\parallel c$
rotates the local moments out of the $ab$-plane without spin-flop.

As shown in Fig.\ \ref{ms001}{\bf a}, at $T =$ 4~K and below $H_{\rm N2}$,
region II, where Eu moments at all three lattice sites are ordered \cite{mor23,
rah23}, the sample contracts with increasing $H$ in an impeccable
$H^2$-fashion. This results from \cite{jan76,bet19} $\Delta{\rm L}(H)\sim
-[M(H)/M_s]^2$ ($M_s$ is the saturation magnetization) and $M(H) \sim H$ in
\begin{figure}[b]
\centering
\includegraphics[width=7.8cm]{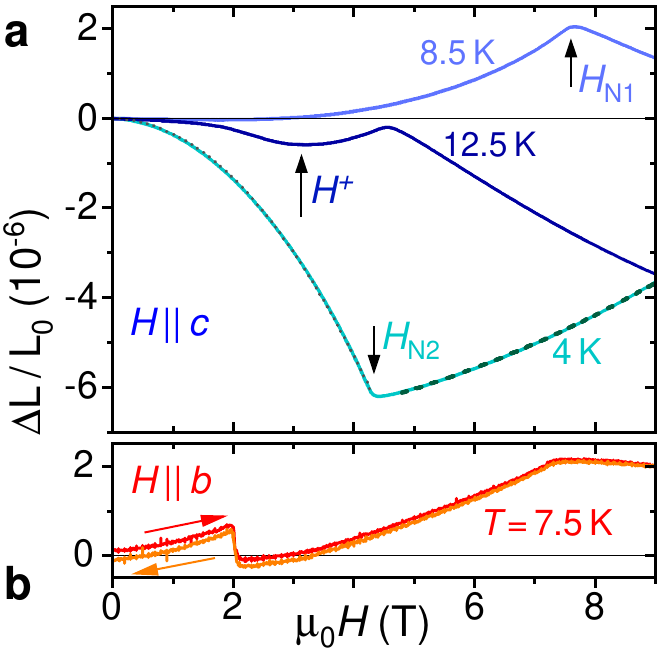}
\caption{\textbf{Magnetostriction data.} {\bf a} $\Delta{\rm L}(H) / $L$_0$
for {\bf H} $\parallel c$ and three exemplary temperatures. The dotted black
line at 4~K is a fit to $\Delta{\rm L}(H) / $L$_0 = a'H^2$ (see text), and the
dashed one to $a'' H^2 + c''$, see Supplementary Note 5. {\bf b}
$\Delta{\rm L}(H) / $L$_0$ for {\bf H} $\parallel b$. Data for field sweep up
(red) and down (orange) show a tiny hysteresis. A spin-flop takes place at
$H_{\rm m}^b \approx$ 2~T.}  \label{ms001}
\end{figure}
this regime \cite{ale23}. In the same phase space, the sample contracts not
only with $H$, but also with increasing $T$, cf.\ Figs.\ \ref{aniso}{\bf a}
and \ref{te-ms}. This is, however, outweighed by a much larger extension
along $a$, both with increasing $T$ and $H$, such that the total sample volume
increases. These effects are reversed upon entering region I for fields above
$H_{\rm N2}$: the sample now expands along $c$ with increasing $H$ or $T$. We
emphasize that the $H^2$-dependence of the magnetostriction inside region I,
dashed line in Fig.\ \ref{ms001}{\bf a}, is consistent with an expected
expansion of fm polarons, see Supplementary Note 5. Unfortunately, with a
maximum field of 9~T along the $c$ axis we do not reach $H_{\rm N1}$ at 4~K.

At 8.5~K, $\Delta{\rm L}(H) / $L$_0$ is negligibly small up to $\sim$2.5~T.
This can be explained by a concurrence of sizeable moment fluctuations at zero
field in this $T$-range, seen e.g.\ by muon spin rotation measurements
($\mu$SR) \cite{rah23}, and a fm alignment of magnetic moments with field, as
will be discussed below. The latter prevails at higher fields and the sample
expands with $H$, as expected for region I from the 4~K-data. Between 5~T
$\lesssim \mu_0 H \lesssim$ 7.4~T, a $H^2$-law holds again. Above $H_{\rm N1}$,
i.e.\ in the paramagnetic regime, fluctuations are suppressed by magnetic field
and the sample contracts.

An intriguing behavior is observed at $T =$ 12.5~K. The sample initially
contracts, but crosses over into expansion at $H^+$ before a sharp transition
into contraction is observed for $H > H_{\rm N1}$. According to these
dilatation data the minimum at $H^+$ indicates a crossover from region III
into I of the phase diagram Fig.\ \ref{phasedia}{\bf c}, in contrast to the
sharp transitions at $H_{\rm N1}$ and $H_{\rm N2}$. Likely, it is this
crossover behavior that hindered a clear-cut assignment in other measurements
\cite{ale23,rah23}. It should be noted that the slopes of
$\Delta{\rm L}^{001}(H) / $L$_0$ just below $H^+$ as well as above $H^+$ are
comparable for the different temperatures.\\

\begin{figure}[t]
\centering
\includegraphics[width=8.2cm]{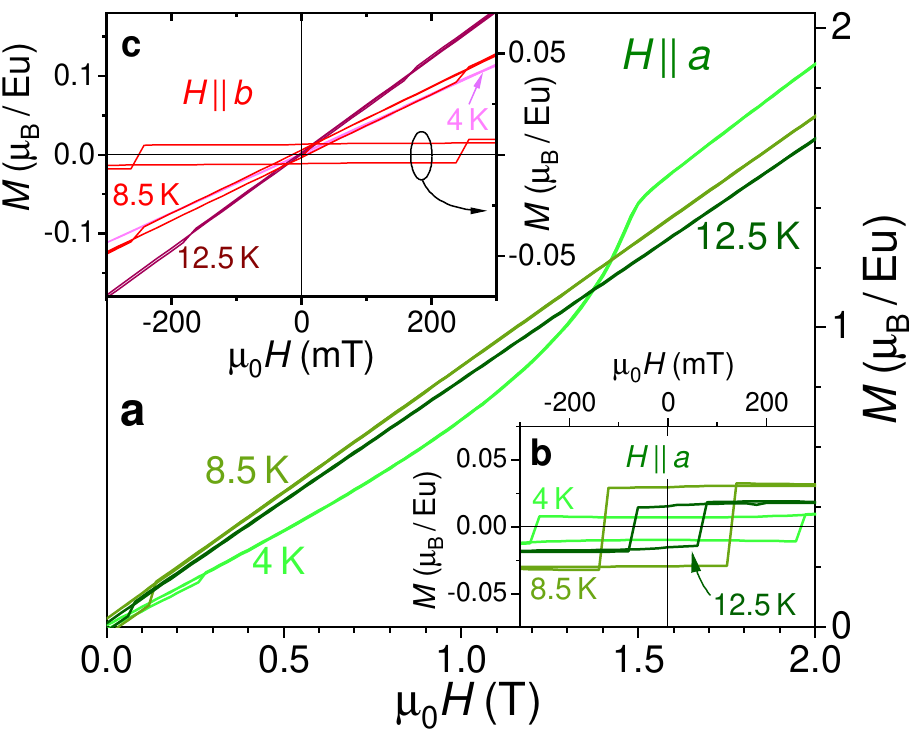}
\caption{\textbf{Magnetization curves $M(H)$ for selected $T$ and $H$.} {\bf a}
At 4~K, a metamagnetic transition is observed for {\bf H} $\parallel a$. {\bf b}
Low-field zoom of data with {\bf H} $\parallel a$ and linear slope subtracted
showing hysteresis. {\bf c} $M(H)$ at low fields for {\bf H} $\parallel b$. A
tiny hysteresis is seen at 8.5~K, as is more obvious once a linear slope is
subtracted, right scale.} \label{MH-ab}
\end{figure}

\noindent \textbf{Thermal expansion at low temperatures}\\
The values of $\alpha^{100}$ and $\alpha^{001}(T, H=0)$ show a qualitatively
similar behavior but with opposite sign, cf.\ Fig.\ \ref{te-ms}{\bf b} and its
inset (see also Supplementary Note 3 and Supplementary Figure 4). Likewise, the
magnetostrictions along $a$ and $c$, Figs.\ \ref{te-ms}{\bf c} and
\ref{ms001}{\bf a}, exhibit qualitatively very similar behavior, yet again with
opposite sign. However, the effects are much larger along $a$ compared to $c$.
In addition, the magnetostriction $\Delta {\rm L}^{100}(H)$ does not follow an
$H^2$-dependence even below $T_{\rm N2}$ as can be inferred from the
coefficient $\lambda$ in Fig.\ \ref{te-ms}{\bf d}. This behavior can be
attributed to the complex magnetic structure of Eu$_5$In$_2$Sb$_6$. The
magnetization $M^{100}(H)$ exhibits a metamagnetic behavior \cite{ale23}; in
case of the example at 4~K this takes place just below 1.5~T, see Fig.\
\ref{MH-ab}{\bf a}. $H_{\rm N2}$ may then be related to a spin-flop transition
within the additional afm order below $T_{\rm N2}$ with ordered moments along
$a$ \cite{mor23,rah23}. The very sharp, almost jump-like increase of $\Delta
{\rm L}^{100}(H)$ near $H_{\rm N2}$ at lowest temperatures supports such a
scenario which, in turn, explains the deviation of $\Delta {\rm L}^{100}(H)$
from a $H^2$-dependence.

Fully relativistic density functional theory (DFT) calculations predict the
ground state of Eu$_5$In$_2$Sb$_6$ to be close to an afm A-type configuration
with the Eu spin moments in the $ab$-plane and antiparallel stacking along $c$
\cite{ale22}. The easy magnetization direction is along $b$ \cite{ros20,ale23}.
Recent neutron data are consistent with this picture, particularly for
$T_{\rm N2} \leq T \leq T_{\rm N1}$ \cite{mor23,rah23}. Given the
aforementioned relation of our dilatation data to the magnetic properties, it
is then not surprising that $\alpha^{010} (T)$ is small, see inset to Fig.\
\ref{aniso}{\bf a}. The magnetostriction at 7.5~K is similarly small, red data
in Fig.\ \ref{ms001}{\bf b}. The magnetic structure facilitates a spin-flop
transition which is clearly seen in $\Delta{\rm L}^{010}(H)$ at $H_{\rm m}^b
\sim$ 2~T. $M(H)$ data revealed this spin-flop to occur at all $T < T_{\rm N1}$
\cite{ale23}. In addition, $\Delta{\rm L}^{010}(H)$ again follows an
$H^2$-dependency up to $H_{\rm N1} \sim$ 7.4~T except near $H_{\rm m}^b$. The
small offset in length after a cycle of field up- and down-sweeps possibly
results from different configurations of magnetic domains \cite{san21}.\\

\noindent {\large \textbf{Discussion}}\\
The thermal expansion coefficients $\alpha^{010}(T,H)$, inset to Fig.\
\ref{aniso}{\bf a}, do not show an additional feature between $T_{\rm N2}
\leq T \leq T_{\rm N1}$ as it was seen along the $a$ and $c$ directions. With
the Eu moments aligned along $b$, this provides a hint toward the origin of the
additional crossover $H^+$ in the $a$- and $c$-phase diagrams. As mentioned,
there is sufficient evidence for polaron formation in Eu$_5$In$_2$Sb$_6$ well
above $T_{\rm N1}$ \cite{ros20,ale22,sou22,gho22,ale23}. However, unlike in fm
materials where the polarons are simply taken up with the onset of magnetic
order \cite{mol01,kam02,das12,poh18,fra21}, in antiferromagnets the polarons
may coexist with the material's global (long-range) order \cite{ume81,kas95,
mol01,mes04,uyu05,kag06}. Upon cooling Eu$_5$In$_2$Sb$_6$ to below
$T_{\rm N1}$, the already existing polarons with their short-range fm
interaction in the $ab$ plane can most easily embed into the afm ordered
environment with their magnetic moments oriented along $b$, as suggested by the
fm chains along $b$, see sketch of moments in Fig.\ \ref{aniso}. Along $b$ and
in region I of the phase diagram, the fm polarons simply include a few reversed
moments around the trapped conduction electron, cf.\ sketch in Fig.\
\ref{phasedia}{\bf a} and Ref.\ \cite{mes04}. Of course, the alignment of
magnetic moments between different fm polarons can still be opposite and hence,
a net antiferromagnetism can be observed along $b$. Note that, despite the afm
ground state, short-range fm interactions are relevant in Eu$_5$In$_2$Sb$_6$ as
evidenced by Curie-Weiss fits of the high-temperature susceptibility
\cite{ros20,ale23}. It is this delicate balance of the two interactions---along
with the gain in exchange energy---which allows fm polarons to easily form
above $T_{\rm N1}$ \cite{jun00,sho19}.

The situation is different for the $a$ direction, Fig.\ \ref{te-ms}. Below
$T_{\rm N1}$, the sample expands with decreasing $T$, a behavior
observed in a few afm materials upon entering the ordered state
\cite{son21,he23}. As a result of the intermediate magnetic anisotropy along
$a$, the magnetization processes of the afm states are different compared to
the $b$ direction. In a simplified, two-sublattice picture of an
antiferromagnet, a magnetic field along $a$ or $c$ results in a canting of the
moments, and polarons with moment parallel {\bf H} will experience a
different change of its surface energy as compared to those antiparallel
moments. In consequence, the effective size of these polarons and their
stability against thermal or quantum mechanical fluctuations will be changed.
(Note that such differences in polaron-moment alignment do not take hold for
field applied in easy $b$ direction.) Such a stabilization is reflected by the
suppression of the negative $\alpha^{100}(T,H)$ just below $T_{\rm N1}$ with
increasing fields up to $\mu_0 H =$ 4~T. Essentially, this picture of embedded
ferromagnetic polarons should also hold for the more complex afm state of
Eu$_5$In$_2$Sb$_6$ in applied fields perpendicular to the easy axis. Hence,
these high-field data suggest that in region I, fm polarons coexist with afm
long-range order. This is corroborated by the negative magnetostriction $\Delta
{\rm L}^{100}$ at higher fields in Fig.\ \ref{te-ms}{\bf c} indicating a
suppression of afm fluctuations. The larger jump of $\alpha^{100}(T,H)$ at
$T_{\rm N1}$ and small fields, i.e.\ upon going from the paramagnetic into
region III in the phase diagram, is then consistent with a prevailing afm
transition at $H_{\rm N1}$ while the crossover at $H^+$ marks the additional
impact of fm polarons. The different afm order below $H_{\rm N2}$ with the
magnetic moments along $a$ \cite{mor23,rah23}, region II, terminates the
signatures of polaron existence as seen by the very strong changes in both
thermal expansion and magnetostriction. The suggestion of a relieved
competition between afm and fm correlations \cite{rah23} is consistent with
this scenario.

The twofold change of slope of the magnetostriction $\Delta{\rm L}
(H)/$L$_0$ in Fig.\ \ref{ms001}{\bf a} clearly indicates competing interactions
and highly likely magnetic inhomogeneity. The resulting phase diagrams exhibit
an unusual $H^+$-crossover line with increasing $H$-values for
increasing $T$ which cannot be explained by anisotropy effects alone
\cite{doe05,bet19}. Rather, such an unexpected thermal behavior is predicted by
a phenomenological model introduced in Supplementary Note 5. This model
considers anisotropic polarons in an antiferromagnet and finds an increased
stability of the fm polarons inside the global afm order upon decreasing
temperature, i.e.\ upon going from region III to I. Concomitantly, the
polarons' volume decreases. As a consequence of these two effects, the
decreasing magnetization of the low-field hysteresis in Figs.\
\ref{MH-ab}{\bf b} and {\bf c}, which indicates a fm polaronic component, can
be rationalized. In contrast, in the more common case of fm polarons in an
eventually fm ordered material, the polarons would just dissolve upon entering
the globally ordered state.

For fields along $a$ and $c$ large enough to drive the system close to the
field-polarized paramagnetic state, polarons with moments opposite to the
field direction cease to exist, leaving only polarons with parallel components
along the field direction. The fact that the $H^+$-crossover lines end on the
$H_{\rm N1}$-lines in the magnetic phase diagrams, Fig.\ \ref{phasedia}, in a
field range comparable to the spin-flop field $H^b_{\rm m}$ for $H \parallel
b$ corroborates such a simple picture. Here, the Zeeman energy compensates the
magnetic anisotropy.

The qualitatively complementary behavior along the $a$ and $c$ axis below
$T_{\rm N1}$, best seen in Fig.\ \ref{te-ms}{\bf b} and its inset, suggests
similar short-range fm and long-range afm competition in the phase diagrams of
these two directions. Hence, the above assignments of the regions I--III can be
cross-compared with the magnetostriction results $\Delta {\rm L}(H)$ for
{\bf H} $\parallel c$ in Fig.\ \ref{ms001}{\bf a}. At 4~K, the initial
decrease of $\Delta {\rm L}(H)$ marks afm region II, while its increase above
$H_{\rm N2}$ in region I (a behavior similar to the one observed before
\cite{hem07}) signals the incorporation of fm polarons. At 8.5~K, long-range
afm and short-range fm correlations coexist and compete, which causes the
initially negligible $\Delta {\rm L}(H)$. Increasing $H$ favors the fm
arrangement and $\Delta {\rm L}(H)$ increases. An increasing volume upon going
from a preferred afm to a fm moment configuration was observed e.g.\ in
(Hf,Ta)Fe$_2$ \cite{li16}. In the paramagnetic regime beyond $H_{\rm N1}$ the
field suppresses magnetic fluctuations and $\Delta {\rm L}(H)$ decreases. At
12~K, the afm order gives rise to the contraction of the sample in region III,
which smoothly crosses over into an expansion above $H^+$, where polarons
appear to exist more easily. The paramagnetic state above $H_{\rm N1}$ is again
accompanied by a negative $\lambda$.

Interestingly, the temperature $T^+$ corresponds to the steepest decrease in
$r(T)$ upon decreasing $T$ (Fig.\ \ref{resist}) consistent with its assignment
as a crossover. A strong decrease of resistivity near 10.5~K in zero
field was also reported earlier \cite{rah23}. We emphasize that the slopes of
$r(T)$ just above $T_{\rm N1}$ and somewhat below $T^+$ coincide, as indicated
by the dotted line in Fig.\ \ref{resist}. This is highly suggestive of polarons
continuing to exist below the onset of afm order at $T_{\rm N1}$ and both
phenomena operating relatively independent of each other.

A physical picture of electronic transport in Eu$_5$In$_2$Sb$_6$ has to
incorporate the existence of self-trapped fm polarons within the afm phases,
which generates a magnetically and electronically inhomogeneous behavior. DFT
results \cite{ale22} suggested Eu$_5$In$_2$Sb$_6$ to be a semimetal with small
Fermi sheets. The availability of charge carriers was found to depend strongly
on the spin configuration, with the fm state having a strongly enhanced density
of states (DOS) at the Fermi level $E_{\rm F}$ compared to (collinear) afm
configurations. The suppression of the fm polaron volume towards lowest $T$
decreases the availability of charge carriers, while the magnetically
inhomogeneous and noncollinear configuration in the ground state impedes
electronic transport. This explains both the strong increase of $\rho(T)$ at
lowest temperatures and the remarkably large magnetoresistance. In
particular, the magnetoresistive properties appear to be similar to those
observed in metallic granular magnetoresistive materials \cite{rao96,ino09}.
An alternative picture arises in insulating or semiconducting materials where
charge carriers (stemming solely from defects) are dressed as fm polarons and
are exclusively responsible for electronic transport. The observed behavior of
$\rho(T)$ upon lowering $T$ should then be explained by the density of polarons
and eventually by increased magnetic scattering of the mobile fm polarons in
the afm matrix. We emphasize that both scenarios rely on the existence of fm
polarons and the self-trapping of charge carriers both in the paramagnetic and
within the afm ordered states. Further investigations of electronic band
structure are needed here.

Our thermal expansion measurements reveal a strongly anisotropic behavior
of Eu$_5$In$_2$Sb$_6$ and nicely confirm the existence of anisotropic polarons
well above $T_{\rm N1}$. In particular, an enhanced DOS at near $E_{\rm F}$
inside the ferromagnetically ordered polarons \cite{ale22} provides an
explanation for the huge negative magnetoresistance in Eu$_5$In$_2$Sb$_6$ which
differs from the double-exchange mechanism in CMR manganites \cite{coe99}. In
the antiferromagnetically ordered phase below $T_{\rm N1}(H)$ the
magnetostriction results point to a coexistence of global afm order with fm
polarons, with the latter being stabilized upon decreasing temperature. In
contrast, there is no indication for polarons below $T_{\rm N2}(H)$. The
insight into polaronic behavior provided here may help in advancing CMR effects
in materials suited for applications.\\

\noindent {\large \textbf{Methods}}\\
Single crystals of Eu$_5$In$_2$Sb$_6$ were grown by a combined In-Sb self-flux
technique \cite{ros20}. Here we report results obtained on the same crystals
as those investigated earlier \cite{ale23} (additional samples investigated
here came from the same batch). The crystallographic orientation of the single
crystals was determined by a real-time Laue X-ray system (Laue-Camera GmbH
\cite{laue}), see also Fig.\ 1{\bf e} in Ref.\ \cite{ale22} and Supplementary
Note 1 and Supplementary Figure 1.

For thermal expansion and magnetostriction measurements a dilatometer cell was
employed \cite{kue12}. The cell was equipped with a Cernox temperature sensor
providing an accuracy of better than $\pm (10\, {\rm mK} + T / 1000)$. The
measurements were conducted in a Physical Property Measurement
System (PPMS) by Quantum Design Inc.\ \cite{qd} with a maximum magnetic field
of 9~T applied parallel to the sample direction along which the dilatation was
measured. Special attention was paid to minimize electrical noise \cite{kue23}.
Thermal expansion measurements were conducted upon warming the sample (if not
stated otherwise) and repeated at least once for comparison. The error of
the measured relative length changes $\Delta$L/L$_0$ in thermal expansion and
magnetostriction was estimated to $\leq 10^{-6}$. The error bar of the
calculated absolute values of the thermal expansion coefficients $\alpha$
amounts to about $5 \cdot 10^{-7}$ K$^{-1}$. This may account for the apparent
low-temperature offsets, particularly along the $b$ direction for which the
sample dimension was smallest (see Supplementary Figure 1). Before each
magnetostriction measurement the sample was heated up to at least 100~K to
avoid any influence of the magnetic history of the sample. Unfortunately, the
samples often broke inside our measurement cell, specifically upon applying
magnetic fields, a fact that severely limited the number of results obtained
at higher fields.

Magnetization measurements were conducted in a Magnetic Property Measurement
System (MPMS3) by Quantum Design Inc.\ \cite{qd} with an accuracy of better than
0.004 $\mu_{\rm B}/$Eu.\\

\appendix
\section{\large Appendix}
\section{\nsize 1. Eu$_5$In$_2$Sb$_6$ samples}
\begin{figure*}[t!]
\centering
\includegraphics[width=14.6cm]{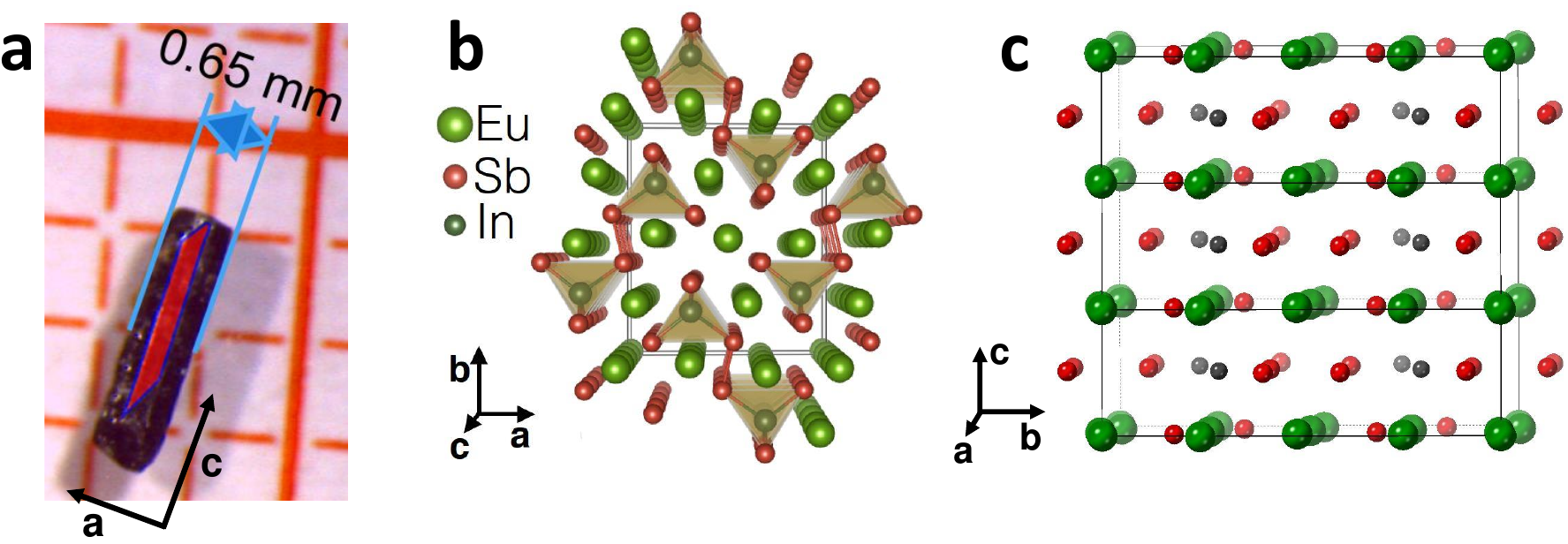}
\caption{\textbf{Eu$_5$In$_2$Sb$_6$ samples.} {\bf a} Photograph of a typical
\eis\ sample used in this study. The crystallographic orientation was
determined by Laue diffraction measurements. All sample are rod-like shaped
with the long dimension corresponding to the crystallographic $c$ direction.
The topmost $ac$ plane of the sample is marked in red. {\bf b} and {\bf c}
Perspective view of the orthorhombic crystal structure of \eis\ along the [001]
and the [100] axis, respectively. In b, the three groups of
crystallographically unique Eu atoms, Eu(1), Eu(2), and Eu(3), are marked.
Figure part b taken from Ref.\ \onlinecite{ale22}.}  \label{sample}
\end{figure*}

Supplementary Fig.\ \ref{sample} shows the dimensions and orientation of one of
our \eis\ samples (sample length along the shortest, $b$ direction is
approximately 0.46~mm) which is typical for the samples investigated in this
study. In line with our earlier findings \cite{ale22}, the $ab$ planes are
usually well defined implying that sample mounting for dilatation measurements
along $c$ is straightforward. However, care must be taken due to the rod-like
shape of the samples which often caused the samples to break during the
experiments. In contrast, the samples did often not exhibit well defined $ac$
and $bc$ planes which rendered sample mounting difficult for dilatation
measurements along $b$ and $a$, respectively. Again, samples often cracked
after measurements leaving the corresponding results unreliable.

\section{\nsize 2. Fitting the thermal expansion data}
In the following, we present detailed results obtained by using different
models to fit our thermal expansion data measured along different
crystallographic directions.\\

\noindent \textbf{Semiclassical model}\\[0.2cm]
For fitting our data to the model developed by Mukherjee \textit{et al.}
\cite{muk94,muk96} we made use of the formulae provided there for the change
of length:
\begin{equation}
\frac{\Delta {\rm L}}{{\rm L}_0} = \frac{\langle \xi \rangle_T - \langle \xi
\rangle_{T_0}}{{\rm L}_{T_0}} \;\; .
\end{equation}
Here, ${\rm L}_{T_0}$ is the length of the sample at reference temperature
$T_0$. The average lattice displacement $\xi$ at a given temperature is
\cite{muk94}:
\begin{equation}
\begin{split}
\langle \xi \rangle_T = \gamma T^2 + \frac{3g}{4c^2} \Bigg[ 3k_{\rm B} T
\left( \frac{T}{\theta_{\rm D}} \right)^3 \int^{\theta_{\rm D}/T}_0
\frac{z^3 dz}{e^z - 1} \\
 - \; \frac{3f}{4c^2} \left( 3k_{\rm B} T \left(
\frac{T} {\theta_{\rm D}} \right)^3 \int^{\theta_{\rm D}/T}_0
\frac{z^3 dz}{e^z - 1} \right)^2 \Bigg]
\end{split}
\end{equation}
For simplicity, we used $u = \frac{3g}{4c^2} 3k_{\rm B}$ and $v = \frac{9gf}
{16c^4} (3k_{\rm B})^2$ when conducting the fits. Also, a vertical offset of
the fitted curves were allowed. The results of the fitting procedure are listed
in the following Table \ref{fitmuk} and some fitted curves are included in
Figure 1 of the main text (dashed lines). $T_{min}$ refers to the minimum
temperature for which data were included in the respective fitting procedure.
The maximum temperature of data used in the fitting was 130~K.\\
\begin{table}[t]
\caption{Fit parameters obtained by fitting $\Delta{\rm L}(T,H)$ to the model
described by Mukherjee \textit{et al.} \cite{muk94}. For simplicity,
$\theta_{\rm D} =$ 197~K was fixed \cite{ale23}.}
\label{fitmuk} \begin{ruledtabular}
\begin{tabular}{c|cc|ccccc}
\multicolumn{3}{c|}{data} & \multicolumn{5}{c}{fit parameter}\\ \hline
data set & direction & $\mu_0 H$ & offset & $\gamma$ & $u$ & $v$ & $T_{min}$ \\
         & & T         & 10$^{-6}$ & $10^{-9}\,$K$^{-2}$ & & & K \\ \hline
\rule{-3pt}{10pt}
36A & (100) &   0 &  4.7    & 36 & 41.3 & 0.62 & 22.8 \\
42B & (100) & 0.5 &  0.5    & 40 & 40.0 & 0.59 & 23.5 \\
40B & (100) & 1.0 & $-3.1$  & 40 & 38.6 & 0.61 & 22.6 \\
41B & (100) & 1.5 & $-16.3$ & 44 & 35.9 & 0.60 & 23.6 \\
39B & (100) & 2.0 & $-30.1$ & 48 & 33.8 & 0.62 & 22.8 \\
38B & (100) & 3.0 & $-16.8$ & 57 & 29.2 & 0.66 & 22.9 \\
37B & (100) & 4.0 & $-0.7$  & 62 & 26.8 & 0.73 & 16.8 \\ \hline
\rule{-3pt}{10pt}
191022 & (001) & 0 & $-0.5$  &  0 & 23.3 & $-0.48$ & 30.5
\end{tabular}
\end{ruledtabular}
\end{table}
\begin{figure*}[t]
\centering
\includegraphics[width=14.8cm]{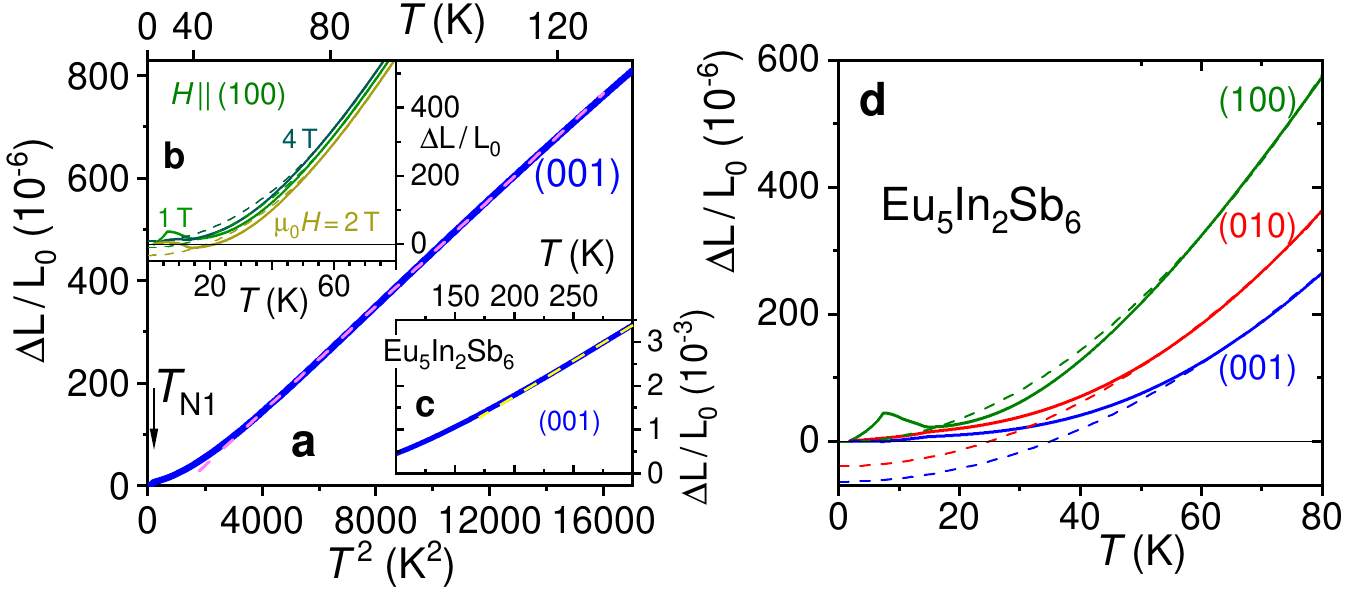}
\caption{\textbf{Fit of the thermal expansion data to parabolas.} {\bf a}
$\Delta{\rm L}^{001}(H) / $L$_0$ plotted on a $T^2$-scale to exemplify the
quality of the fit up to 130~K. {\bf b} $T^2$-fits to data obtained at applied
magnetic fields. {\bf c} At $T \gtrsim$ 175~K a linear expansion is found (as
discussed in the main text). {\bf d} Examples of parabolic fits with focus on
temperatures below 80~K. The inset shows exemplary fits to data obtained at
finite magnetic fields along (100). The parameters obtained from the fits are
summarized in Supplementary Table \ref{fitpara}.  The error of the measured
relative length changes $\Delta$L/L$_0$ was estimated to $\leq 10^{-6}$.}
\label{fitT2}
\end{figure*}

\noindent \textbf{Parabolic fits}\\
For simplicity, we also attempted to fit our thermal expansion data to a
parabolic function $\Delta{\rm L}(T) /$L$_0 = c + aT^2$. The resulting
fitting parameters are listed in Table \ref{fitpara}. The fitted curves are
presented in Fig.\ \ref{fitT2}.
\begin{table}[!h]
\caption{Fit parameters obtained by fitting the thermal expansion data to
$\Delta{\rm L}(T) /$L$_0 = c + aT^2$. Here, $\Delta{\rm L}(T)$ was measured
for different directions and magnetic fields. $T_{min}$ denotes the minimum
temperature of data included in the fit.}  \label{fitpara}
\begin{ruledtabular}
\begin{tabular}{c|cc|ccc}
\multicolumn{3}{c|}{data} & \multicolumn{3}{c}{fit parameter}\\ \hline
data set & direction & $\mu_0 H$ & $c$       & $a$      & $T_{min}$ \\
         &           & T         & 10$^{-6}$ & K$^{-2}$ & K  \\ \hline
\rule{-3pt}{10pt}
36A & (100) &   0 &  $-5.9$ & 0.0905   & 55.0 \\
42B & (100) & 0.5 &  $-7.5$ & 0.091    & 55.0 \\
40B & (100) & 1.0 & $-10.9$ & 0.0915   & 55.6 \\
41B & (100) & 1.5 & $-21.9$ & 0.0912   & 54.5 \\
39B & (100) & 2.0 & $-33.2$ & 0.0913   & 54.5 \\
38B & (100) & 3.0 & $-13.3$ & 0.0915   & 55.0 \\
37B & (100) & 4.0 &  $+8.5$ & 0.0918   & 55.0 \\ \hline
\rule{-3pt}{10pt}
241022 & (010) &   0 & $-39.3$ & 0.0626   & 48.0 \\
191022 & (001) &   0 & $-63.8$ & 0.0517   & 58.0 \\
\end{tabular}
\end{ruledtabular}
\end{table}

\begin{figure}[h]
\centering
\includegraphics[width=8.2cm]{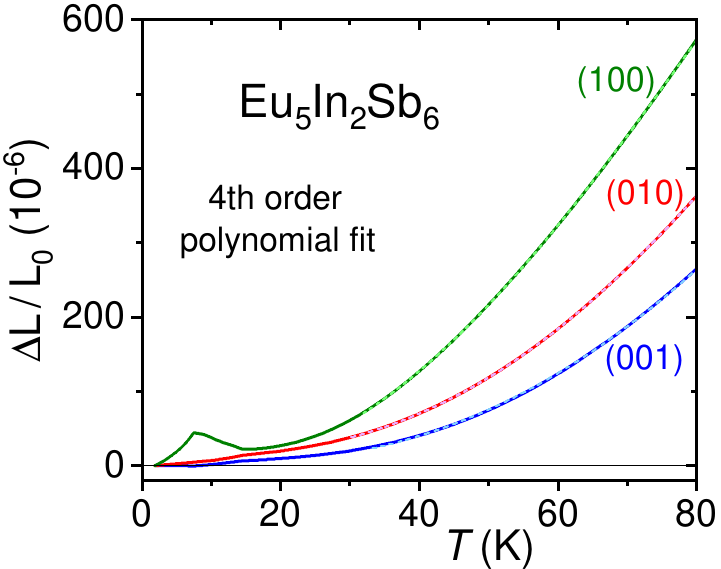}
\caption{\textbf{Fit of the thermal expansion data to polynomials of fourth
order.} Fitted curves are shown as light dashed lines. The parameters obtained
from the fits are summarized in Supplementary Table \ref{fitpoly}. The error
of $\Delta$L/L$_0$ is $\leq 10^{-6}$.} \label{fitT4}
\end{figure}

\noindent \textbf{Fits of polynomials of 4th order}\\
In addition, the thermal expansion data are fitted by using polynomials of
fourth order. The results of these fits are presented in Table \ref{fitpoly}.
Note that linear terms have been omitted. The fitted curves are shown in Fig.\
\ref{fitT4}.
\begin{table}[!h]
\caption{Fit parameters obtained by fitting the thermal expansion data at zero
field to $\Delta{\rm L}(T) /$L$_0 = \sum_{i=0}^4 a_i T^i$. $T_{min}$ denotes
the minimum temperature of data included in the fit. The linear coefficient was
set to $a_1 = 0$.}  \label{fitpoly}
\begin{ruledtabular}
\begin{tabular}{c|c|ccccc}
\multicolumn{2}{c|}{data} & \multicolumn{5}{c}{fit parameter}\\ \hline
data set & direction & $a_0$   & $a_2$ & $a_3$ & $a_4$  & $T_{min}$ \\
 & & $10^{-6}$ & $10^{-9}$ & $10^{-10}$ & $10^{-12}$ & \\
 & & & K$^{-2}$ & K$^{-3}$ & K$^{-4}$  & K  \\ \hline \rule{-3pt}{10pt}
36A      & (100)     & $-22.4$ & 72.4  & 8.1   & $-7.0$ & 32.0 \\
241022   & (010)     &  10.8   & 3.2   & 10.6  & $-5.1$ & 30.2 \\
191022   & (001)     & $-1.7$  & 1.6   & 7.4   & $-3.0$ & 33.4 \\
\end{tabular}
\end{ruledtabular}
\end{table}

\section{\nsize 3. Thermal expansion data measured along
the $b$ and $c$ crystallographic direction}
\begin{figure}[h]
\centering
\includegraphics[width=7.6cm]{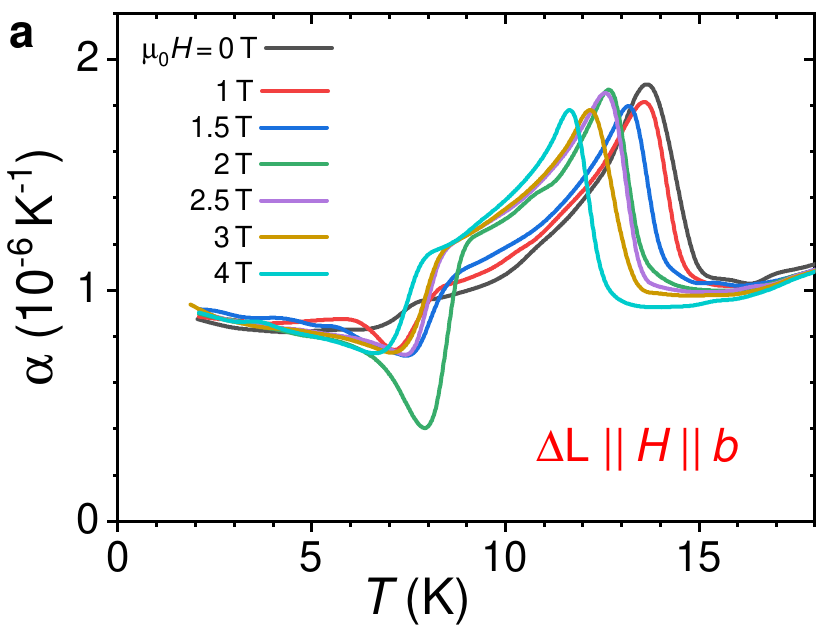}
\includegraphics[width=7.6cm]{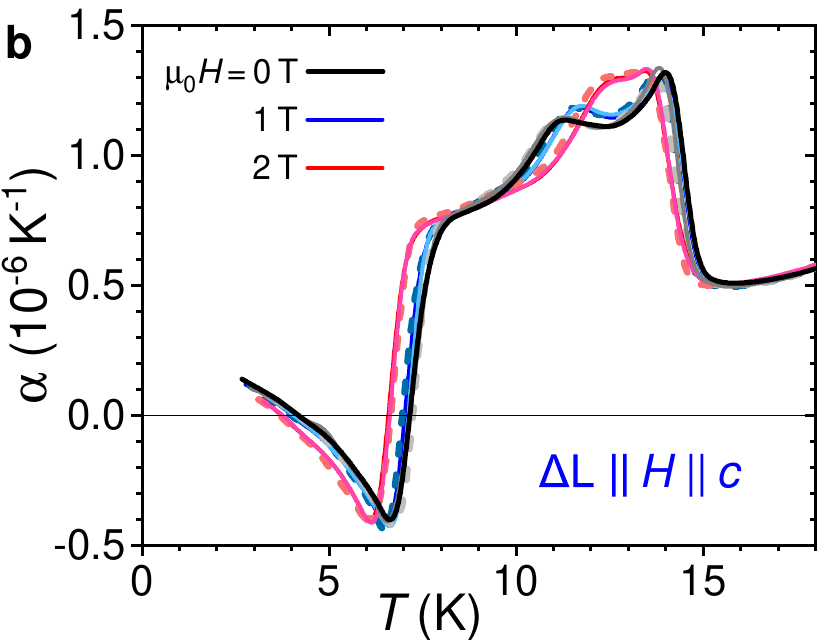}
\caption{\textbf{Thermal expansion coefficients $\alpha$} measured along
{\bf a} the crystallographic $b$ direction and {\bf b} the $c$ direction. In
{\bf b}, results of three consecutive runs for each field value (0 T, 1 T, 2 T)
are presented by lines of similar colors; full lines: sample warming up to 100
K, dashed lines: sample cooling down. Errors bars of $\alpha$ amount to
about $5 \cdot 10^{-7}$ K$^{-1}$.}
\label{alpha-bc} \end{figure}

Some thermal expansion data measured along the $b$ and $c$ crystallographic
directions are shown in Fig.\ \ref{alpha-bc}.

\section{\nsize 4. Additional magnetostriction data}
\begin{figure}[t]
\centering
\includegraphics[width=7.8cm]{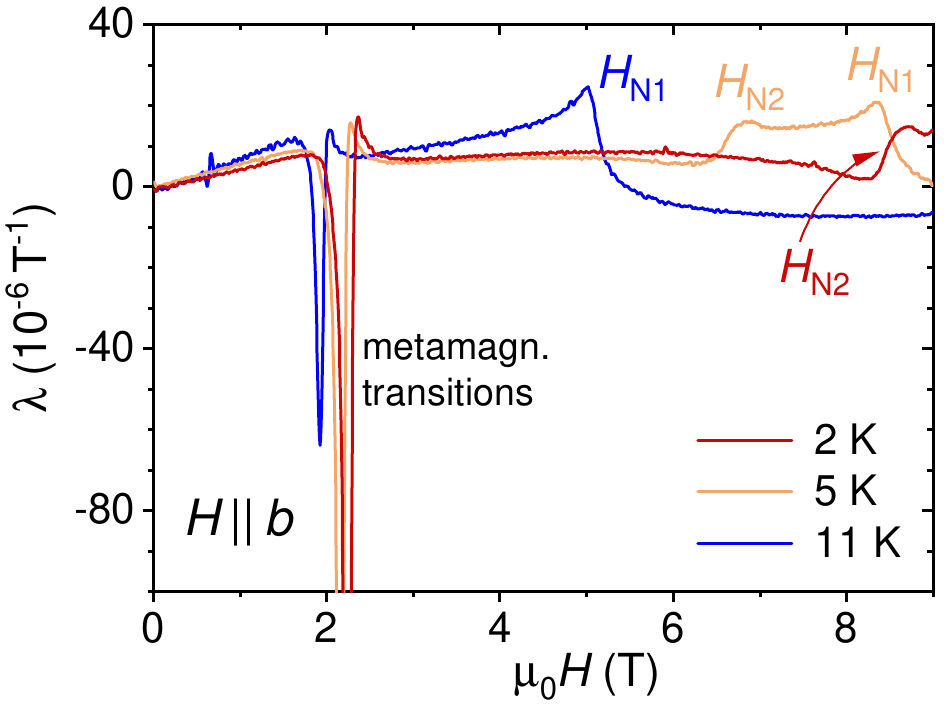}
\caption{\textbf{Magnetostriction coefficient $\lambda$} measured with {\bf H}
$\parallel b$ at three different temperatures.}  \label{ms010PR}
\end{figure}

In an effort to increase reliability, the thermal expansion and
magnetostriction measurements were conducted on several samples and by using
two different measurement setups. In addition to the data shown in the main
text, we report in Fig.\ \ref{ms010PR} magnetostriction data obtained by means
of another setup for $\vec{H} \parallel b$. The corresponding results are 
included in the phase diagram Figure 5{\bf b} of the main text as dotted 
triangles. Clearly, these data fit excellently into the phase diagram.

Thermal expansion and magnetostriction measurements shown in this Note were
performed using a capacitance dilatometer in a $^4$He cryostat\cite{sch06}.
All thermal expansion measurements were performed using a slow continuous
temperature ramp, whereas all magnetostriction measurements were
performed by stabilizing the field to avoid the influence of eddy currents.
Thermal expansion data were corrected by performing a background subtraction
of the cell effect under identical thermal conditions.

\section{\nsize 5. Description of anisotropic polarons}

\noindent \textbf{Anisotropic polarons in an anisotropic antiferromagnet}\\
Hereinafter, we closely follow the procedure of Kagan \textit{et al.}
\cite{kag06}. Due to the anisotropy in the exchange coupling and the hopping of
the conduction electrons the polarons are assumed to have ellipsoidal shape
with axes $\{R_x,R_y,R_z\}$.  We consider the Hamiltonian
\begin{equation}
\begin{split}
{\cal {\hat H}} = - J_H \sum_i {\vec S}_i \cdot {\vec \sigma}_i + \sum_{\langle
ij \rangle_{\alpha}} J_{\alpha} \Bigl({\vec S}_i \cdot {\vec S}_j + S^2\Bigr)
\\ - \sum_{\langle ij \rangle_{\alpha} \sigma} t_{\alpha}
\Bigl(c^{\dagger}_{i \sigma} c_{j \sigma} + H.c. \Bigr)
\end{split}
\end{equation}
on a simple cubic lattice structure of the antiferromagnetic material.  Here,
$J_H$ is the Hund's rule interaction which locally couples ferromagnetically
the spin ${\vec S}$ with the spin of the electron ${\vec \sigma}$, $J_{\alpha}$
is the nearest neighbor exchange coupling in the direction $\alpha =\{x,y,z\}$,
and $t_{\alpha}$ the nearest neighbor hopping amplitude along the
$\alpha$-direction. We assume $J_H \gg \{t_x,t_y,t_z\} \gg\{J_x,J_y,J_z\}$ and
denote the intersite distance with $a$. The exchange energy has been shifted
such that the antiferromagnetic state has zero energy. The spins are
semiclassical, ${\vec S} = 2S {\vec \sigma}$, i.e.\ they correspond to a spin
1/2 with the length of a spin $S$.

A prerequisite for polaron formation is a low enough carrier concentration. The
dispersion of the conduction band is small, so that the electrons are spread
over the entire polaron. We consider two interpenetrating sublattices, A and B,
with up- and down-spins, respectively. Due to the Hund's rule coupling the
sublattices provide up-spin and down-spin charge carriers, respectively. A very
small magnetic field polarizes the conduction band.

An ellipsoidal polaron has a dimensionless volume of $\Omega_{\rm ell} =
\frac{4\pi}{3} \frac{R_xR_yR_z}{a^3}$ and a polarization energy of
\begin{eqnarray}
E_{\rm pol} & = & 2(J_x+J_y+J_z)S^2n\Omega_{\rm ell} \nonumber \\
 & = & 2(J_x+J_y+J_z)S^2n (4\pi/3) \frac{R_xR_yR_z}{a^3} \ ,
\end{eqnarray}
that arises from the exchange energy of $2J_{\alpha}S^2$ per bond inside the
ellipsoid, while the exchange energy at the surface boundary vanishes. The
conduction electrons inside the polaron have carrier density $n$ and their
spins aligned either all up or all down.

The kinetic energy of the band of conduction states with nearest neighbor
hopping is
\begin{equation}
\varepsilon({\vec k}) = -2\Bigl[t_x\cos(k_xa) + t_y\cos(k_ya) + t_z\cos(k_za)
\Bigr] \ .
\end{equation}
Expanding for small ${\vec k}$ we have
\begin{eqnarray}
\varepsilon({\vec k}) & = &-2(t_x+t_y+t_z) + [t_x(k_xa)^2+t_y(k_ya)^2+
t_z(k_za)^2] \nonumber \\
& \to & -2(t_x+t_y+t_z) - a^2[t_x\frac{\partial^2}{\partial x^2}+
t_y\frac{\partial^2}{\partial y^2} +t_z\frac{\partial^2}{\partial z^2}] .
\end{eqnarray}
For simplicity, we rescale the coordinates ${\tilde y} = y \sqrt{\frac{t_x}
{t_y}}$ and ${\tilde z} = z \sqrt{\frac{t_x}{t_y}}$ so that
\begin{equation}
R_x = {\tilde R}_x = {\tilde R}_y = R_y \sqrt{\frac{t_x}{t_y}} = {\tilde R}_z =
R_z \sqrt{\frac{t_x}{t_z}} \nonumber
\end{equation}\
and
\begin{equation}
\nabla_{\tilde r}^2 = \frac{\partial^2}{{\partial {\tilde r}}^2} +
\frac{2 \partial}{{\tilde r} \partial{\tilde r}}
\end{equation}
Without loss of generality, we assumed $t_x > \{t_y,t_z\}$ resulting in
$R_x$ being the largest axis of ellipsoid and
\begin{equation}
{\cal {\hat H}}_{\rm kin} = -a^2t_x [\frac{\partial^2}{\partial
x^2}+\frac{\partial^2}{\partial {\tilde y}^2}+\frac{\partial^2}{\partial
{\tilde z}^2}] = -a^2t_x \nabla_{\tilde r}^2 \;\; .
\end{equation}
The Schr\"odinger equation is
\begin{equation}
{\cal {\hat H}}_{\rm kin} \psi({\vec {\tilde r}}) = - a^2t_x \nabla_{\tilde
r}^2 \psi({\vec {\tilde r}}) = \varepsilon_0 \psi({\vec {\tilde r}}) ,
\end{equation}
where $\varepsilon_0$ is the ground state energy. The ground state wave
function in the polaron $\psi({\vec {\tilde r}}) = \frac{\sin(k{\tilde r})}
{k{\tilde r}} = j_0(k{\tilde r})$ should vanish at the surface of the ellipsoid
so that $k = \pi/{\tilde R}$. The ground state energy is then
\begin{equation}
\varepsilon_0 = t_xa^2(\pi/R_x)^2 \ .
\end{equation}
We assume that the charge density is dilute enough so that all the electrons
can be accommodated into the ground state.  The kinetic energy and the total
energy are then given by
\begin{eqnarray}
E_{\rm kin} & = & -2(t_x+t_y+t_z)n + t_xa^2(\pi/R_x)^2n \\
E_{\rm tot} & = & E_{\rm pol} + E_{\rm kin} = -2(t_x+t_y+t_z)n + t_xa^2
(\pi/R_x)^2 n \nonumber \\
& & + 2(J_x+J_y+J_z)S^2 n (4\pi/3) \left(\frac{R_x}{a}\right)^3
\frac{\sqrt{t_yt_z}}{t_x} .
\end{eqnarray}
With the notations ${\tilde J} = (J_x+J_y+J_z)S^2$ and $t_{\rm eff} =
(t_xt_yt_z)^{1/3}$ we find
\begin{equation}
\begin{split}
E_{\rm tot} = -2(t_x+t_y+t_z)n + t_x\pi^2(a/R_x)^2n \\
+\, 2{\tilde{J}}n (4\pi/3) \left(\frac{R_x}{a}\right)^3
\left(\frac{t_{\rm eff}} {t_x}\right)^{3/2} \ .
\end{split}
\end{equation}
The minimization of the total energy with respect to $R_x/a$ yields for the
large axis of the ellipsoid
\begin{eqnarray}
R_x/a & = & \left(\frac{\pi t_x^2}{4 {\tilde J}\sqrt{t_yt_z}}\right)^{1/5} =
\left[\frac{t_x \pi}{4{\tilde J}}\left(\frac{t_x}{t_{\rm eff}}\right)^{3/2}
\right]^{1/5} \\
\Omega_{\rm ell} & = & \frac{4 \pi}{3} \frac{R_xR_yR_z}{a^3} = \frac{4 \pi}{3}
\left( \frac{R_x}{a}\right)^3 \sqrt{\frac{t_yt_z}{t_x^2}}  \\ \nonumber
 & = & \frac{4 \pi}{3} \sqrt{\frac{t_yt_z}{t_x^2}} \left(\frac{\pi t_x^2}
{4{\tilde J} \sqrt{t_yt_z}} \right)^{3/5} = \frac{\pi^{8/5} 2^{4/5}}{3}
\left( \frac{t_{\rm eff}}{\tilde J} \right)^{3/5}
\end{eqnarray}
Hence, the energy of an ellipsoidally shaped polaron amounts to
\begin{eqnarray}
E_{\rm tot} & = & -nJ_H S -2(t_x+t_y+t_z)n + t_x\pi^2(a/R_x)^2n \nonumber \\
& & + \, 2 {\tilde{J}}n (4\pi/3)\left(\frac{R_x}{a}\right)^3
\left(\frac{t_{\rm eff}}{t_x}\right)^{3/2}.  \label{etote}
\end{eqnarray}
All terms of the total energy are proportional to $n$. These results are in
agreement with those by Kagan {\it et al}. \cite{kag06}.\\

\noindent \textbf{Anisotropic polarons in a magnetic field (Zeeman
splitting)}\\
In the following, we assume that the magnetic field is applied either parallel
or antiparallel to the ferromagnetic moment of the elliptical polaron giving
rise to opposite signs of the Zeeman energy of the charge carriers
\begin{equation}
E_{\rm kin} = -2(t_x+t_y+t_z)n + t_x \pi^2(a/R_x)^2n \mp g_e \frac{H}{2}n \ .
\end{equation}

We now consider the polarization of the antiferromagnetic lattice, which,
again, is subdivided in sublattices A and B with up-spins and down-spin,
respectively. The ground state of the antiferromagnetic (afm) phase yields
$S_{\rm eff} = 0$. The Zeeman splitting energy is given by $+g_SH \sum_j
S_j^z$, where the sum is over the sublattices A and B. There is a continuum of
magnetic excitations above the ground state singlet. In the afm bulk material,
the additional (field-dependent) term of the energy is then ${\textstyle
\frac{1}{2}} H^2 \chi_{\parallel,\perp}$, where the susceptibility $\chi
\propto 1/{\tilde J}$. This term is small and can be neglected.

The magnetic energy of the ellipsoidal polaron is $\pm g_SSH \Omega_{\rm ell}$
and it follows (proceeding as above) \cite{emi88}
\begin{eqnarray}
E_{\rm pol}^{\pm} &=& (4\pi/3)\left(\frac{R_x}{a}\right)^3 \left(
\frac{t_{\rm eff}}{t_x}\right)^{3/2} \left(2{\tilde{J}}n \mp g_SSH\right)
\nonumber \\
E_{\rm tot}^{\pm} &=& -2(t_x+t_y+t_z)n + t_x \pi^2(a/R_x)^2n \mp g_e \frac{H}
{2}n \nonumber \\
& & + (4\pi/3)\left(\frac{R_x}{a}\right)^3 \!\! \left(\frac{t_{\rm eff}}
{t_x}\right)^{3/2} \!\! \left(2{\tilde{J}}n \mp g_S SH\right) \! .
\end{eqnarray}
The minimization of the total energy with respect to $R_x/a$ yields the radius
$R_x/a$
\begin{eqnarray}
\frac{\partial E_{\rm tot}^{\pm}}{\partial (R_x/a)} & = & 0 =
-2t_x\pi^2(a/R_x)^3n + 4\pi(R_x/a)^2 \left(\frac{t_{\rm eff}}{t_x}\right)^{3/2}
\!\! \zeta \nonumber \\
R^{\pm}_x/a & = &\left[\frac{\pi t_x}{4{\tilde J}}\left(\frac{t_x}{t_{\rm eff}}
\right)^{3/2}\right]^{1/5} \left(\frac{\zeta}{2{\tilde J}n}\right)^{-1/5}
\label{Rxa} \\
\Omega^{\pm}_{\rm ell} & = & \frac{4 \pi}{3} \frac{R_xR_yR_z}{a^3} \nonumber \\
& = & \frac{4 \pi}{3} \sqrt{\frac{t_yt_z}{t_x^2}} \left(\frac{\pi t_x^2}
{4{\tilde J} \sqrt{t_yt_z}}\right)^{3/5} \left(\frac{\zeta}{2{\tilde J}n}
\right)^{-3/5} \nonumber \\
 & = & \frac{\pi^{8/5} 2^{4/5}}{3} \left( \frac{2 n t_{\rm eff}}{\zeta}
 \right)^{3/5}
\end{eqnarray}
where $\zeta = 2{\tilde{J}}n \mp g_S SH$. This way we obtain the total energy
of a polaron in magnetic field
\begin{eqnarray}
E_{\rm tot}^{\pm} & = & -nJ_HS -2(t_x+t_y+t_z)n \nonumber \\
& & \mp \, g_e\frac{H}{2}n + 2^{4/5} \frac{5}{3} \pi^{8/5} t_{\rm eff}^{3/5} n
\left(\frac{\zeta} {2 n}\right)^{2/5} \label{esph}
\end{eqnarray}
from which the magnetization and susceptibility of the polaron is calculated
\begin{align}
m^{\pm} & = -\frac{\partial E_{\rm tot}^{\pm}}{\partial H} \nonumber \\
 & = \pm g_e\frac{n}{2}
\pm 2^{4/5} \frac{1}{3} \pi^{8/5} t_{\rm eff}^{3/5} {\tilde J}^{-3/5} g_SS
\left( \frac{\zeta}{2{\tilde J}n}\right)^{-3/5} \label{msph} \\
\chi^{\pm} & = \frac{\partial m^{\pm}}{\partial H} = 2^{-1/5} \frac{1}{5}
\pi^{8/5} t_{\rm eff}^{3/5} {\tilde J}^{-8/5} \frac{(g_SS)^2}{n} \left(
\frac{\zeta} {2{\tilde J}n}\right)^{-8/5} . \label{xsph}
\end{align}

Polarons with their spins aligned parallel to the magnetic field, $E^+$, have
lower energies, are of larger size and have larger magnetization compared to
those with antiparallel alignment, $E^-$, as well as $\chi^+ > \chi^-$.  Note
that $\Omega^+_{\rm ell}$, $m^+$ and $\chi^+$ diverge when $g_SSH \to 2{\tilde
J}n$ and the entire sample is field-polarized. For fields smaller than the one
required for sample polarization, at finite temperature, a mixture of $E^+$ and
$E^-$ polarons with opposite magnetization is present in the material.

It is important to note that a softening of the antiferromagnet, e.g.\ by
increasing the temperature or the magnetic field, results in a reduction of
${\tilde J}$. Consequently, $\Omega_{\rm ell}$ increases with $T$ while the
polaron's energy decreases. The polarons grow in size but become less stable
with increasing $T$.  Hence, the magnetization and the susceptibility increase
as well.

We also note that the polaron's zero-field energy is proportional to the
carrier density $n$, while its zero-field magnetization (neglecting $g_en/2$)
is independent of $n$. On the other hand, the susceptibility at $H=0$ is
inversely proportional to $n$.\\

\noindent \textbf{Magnetostriction of polarons}\\
Within the present model the magnetostriction is isotropic in magnetic field,
\begin{equation}
\textstyle\frac{R_x^{\pm}(H)-R_x^{\pm}(0)}{R_x^{\pm}(0)} =
\frac{R_y^{\pm}(H)-R_y^{\pm}(0)}{R_y^{\pm}(0)} =
\frac{R_z^{\pm}(H)-R_z^{\pm}(0)}{R_z^{\pm}(0)} \ . \nonumber
\end{equation}
At $T=0$, all the polarons are of the $E^+$-type and, importantly, they expand
with field. At finite temperature both types of polarons populate the system
and the magnetostriction is given by
\begin{equation}
\textstyle \frac{\Delta{\rm L}(H)} {{\rm L}_0} \propto
\frac{N^+}{N} \frac{R_x^+(H)-R_x^+(0)}{R_x^+(0)} + \frac{N^-}{N} \frac{R_x^-(H)
-R_x^-(0)}{R_x^-(0)} \ , \label{stric}
\end{equation}
where $N^+$ and $N^-$ are the number of polarons of the $E^+$- and $E^-$-type,
and $N=N^++N^-$ is the total number of polarons.

The polarons of $E^-$-type compress under an applied magnetic field, but less
than the expansion of the $E^+$-polarons. In particular, an expansion of eq.\
(\ref{stric}) by using eq.\ (\ref{Rxa}) for small enough magnetic fields,
$g_SSH / 2 {\tilde J}n \ll 1$, results in an $H^2$-dependence of the
magnetostriction to lowest order because linear-in-$H$ terms cancel out. Our
experimental data at $T =$ 4~K and for $\vec{H} \parallel c$ are consistent
with such a description, see Fig.\ 6 of the main text: The data can nicely be
fitted by $\Delta{\rm L}^{001}(H) / $L$_0 = a'' H^2 + c''$ within the field
range 4.8~T $\leq \mu_0 H \leq$ 9~T, where $a'' = 4.25 \cdot 10^{-8}$ T$^{-2}$
and $c'' = -7.1 \cdot 10^{-6}$.\\

\noindent \textbf{Thermal equilibrium}\\
We introduce Boltzmann factors for $N^{\pm}$ and $E^{\pm}$ and the partition
function \cite{emi88}
\begin{equation}
\frac{N^{\pm}}{N} = e^{-E^{\pm}/T} \ \ {\rm and} \ \
{\cal{Z}} = e^{-E^+/T}+e^{-E^-/T} \ .
\end{equation}
Making use of the definitions $\Delta = E^- - E^+$ and $E^0 = (E^- + E^+)/2$
we have
\begin{eqnarray*}
E^{\mp} & = & E^0 \pm \Delta/2 \\
{\cal{Z}} & = & e^{-E^0/T} \, 2 \cosh(\Delta/2T) \\
\frac{N^+-N^-}{N} & = & \tanh(\Delta/2T) \ ,
\end{eqnarray*}
and the free energy is then given by
\begin{eqnarray}
{\cal F} & = & -T \ln({\cal Z}) = E^0 -T \ln[2\cosh(\Delta/2T)] \nonumber \\
 & \approx & E^0 - T \ln(2) - T\ln\left[ 1 + \frac{\Delta^2}{8T^2}\right]
 \nonumber \\  & \approx & E^0 - T \ln(2) - \frac{\Delta^2}{8T} \ .
\end{eqnarray}
The magnetization and susceptibility are then determined as
\begin{eqnarray}
m & = & - \textstyle\frac{\partial {\cal F}}{\partial H} = \frac{\Delta}{4T}
\frac{\partial \Delta} {\partial H} \\
\chi & = & \textstyle\frac{1}{4T} \left( \frac{\partial \Delta}{\partial H}
\right)^2
\end{eqnarray}
For $H \to 0$, the magnetization $m$ vanishes and $\chi$ obeys a Curie law.
In order to obtain a Curie-Weiss law, the polarons would have to interact,
e.g.\ via an exchange interaction, or overlap.

\vspace{0.8cm}\noindent {\large \textbf{Acknowledgements}}\\
S.W. acknowledges fruitful discussion with Oliver Stockert. Work at the
Max-Planck-Institute for Chemical Physics of Solids in Dresden and at Goethe
University Frankfurt was supported by the Deutsche Forschungsgemeinschaft (DFG,
German Research Foundation), Project No.\ 449866704. Work at Los Alamos was
performed under the auspices of the U.S.\ Department of Energy, Office of Basic
Energy Sciences, Division of Materials Science and Engineering. M.S.C.\
acknowledges support from the Los Alamos Laboratory Directed Research and
Development Program.\\

\noindent {\large \textbf{Author contributions}}\\
S.W.\ and J.M.\ conceived the experiments, M.S.C.\ and P.F.S.R.\ prepared the
samples, H.D.-D., M.V.A.C.\ and S.M.T.\ conducted the experiments, U.K.R.\ and
P.S.\ provided theoretical insight, M.V.A.C.\ and S.W.\ analysed the results,
S.W.\ and P.S.\ wrote the draft, with help from M.V.A.C., J.M.\ and U.K.R. All
authors contributed to the discussion of the results and the revision of the
manuscript.\\

\noindent {\large \textbf{Competing interests}}\\
The authors declare no competing interests.
\end{document}